\documentclass[12pt,preprint]{aastex}

\slugcomment{Accepted for publication in the Astrophysical Journal}

\usepackage{ulem}
\usepackage{color}
\usepackage{url}

\shorttitle{DISK FRACTION IN A LOW-METALLICITY ENVIRONMENT}

\shortauthors{Yasui et al.}

\begin{document}

\title{THE LIFETIME OF PROTOPLANETARY DISKS IN A LOW-METALLICITY
ENVIRONMENT\footnote{Based on data collected at Subaru Telescope, which
is operated by the National Astronomical Observatory of Japan.}}

\author{Chikako Yasui and Naoto Kobayashi}
\affil{Institute of Astronomy, School of Science, University of Tokyo,
2-21-1 Osawa, Mitaka, Tokyo 181-0015, Japan}
\email{ck$_-$yasui@ioa.s.u-tokyo.ac.jp}

\author{Alan T. Tokunaga}
\affil{Institute for Astronomy, University of Hawaii, 2680 Woodlawn
Drive, Honolulu, HI 96822, USA}

\author{Masao Saito} 
\affil{ALMA Project, National Astronomical Observatory of Japan, 2-21-1
Osawa, Mitaka, Tokyo 181-8588, Japan}

\and

\author{Chihiro Tokoku} 
\affil{Astronomical Institute, Tohoku University,
Aramaki, Aoba, Sendai 980-8578, Japan}


\begin{abstract}

The extreme outer Galaxy (EOG), the region with a Galactic radius of
more than 18\,kpc, is known to have very low metallicity, about
one-tenth that of the solar neighborhood.  We obtained deep
near-infrared (NIR) images of two very young ($\sim$0.5\,Myr)
star-forming clusters that are one of the most distant embedded clusters
in the EOG.  We find that in both clusters the fraction of stars with
NIR excess, which originates from the circumstellar dust disk at radii
of $\leq$0.1\,AU, is significantly lower than those in the solar
neighborhood.  Our results suggest that most stars forming in the
low-metallicity environment experience disk dispersal at an earlier
stage ($<$1\,Myr) than those forming in the solar metallicity
environment (as much as $\sim$5--6\,Myr).  Such rapid disk dispersal may
make the formation of planets difficult, and the shorter disk lifetime
with lower metallicity could contribute to the strong metallicity
dependence of the well-known ``planet-metallicity correlation'', which
states the probability of a star hosting a planet increases steeply with
stellar metallicity.  The reason for the rapid disk dispersal could be
increase of the mass accretion rate and/or the effective far-ultraviolet
photoevaporation due to the low extinction; however, another unknown
mechanism for the EOG environment could be contributing significantly.

\end{abstract}



\section{INTRODUCTION}
\label{sec:intro}

Because the EOG is known to have an extreme environment such as very low
metallicity, [O/H]$\sim$$-1$ \citep{Rudolph2006}, it can be studied as a
``laboratory'' for the star formation process in an environment totally
different from the solar neighborhood \citep{Ferguson1998,NK2008}. As a
part of our studies of star formation in the EOG
\citep{My2006ApJ,My2008ApJ}, we explore in this paper the lifetime of
circumstellar disks in a low-metallicity environment.

The frequency of disks harboring stars within a young embedded cluster,
the disk fraction, has been studied for many clusters of various ages as
an indicator of the disk lifetime \citep{Haisch2001ApJL,Hernandez2007}.
The most common means of estimating the disk fraction is the detection
of infrared emission from warm and hot dust (typically a few 100 to
1500\,K) in the inner part of the disk (within a stellocentric distance
of $\sim$1\,AU) \citep{Kenyon1987} with wavelengths of 2--24$\,\mu$m
using $K$-band (2.2$\,\mu$m) \citep{Lada1999,Hillenbrand2005}, $L$-band
(3.5$\,\mu$m) \citep{Haisch2001ApJL}, and {\it Spitzer IRAC}
3.6--8$\,\mu$m and/or {\it MIPS} 24$\,\mu$m \citep{DFSpitzer} filters.
The disk fraction was found to decrease as a function of age, showing
that the inner disk lifetime is about 5--10\,Myr for nearby embedded
clusters with the solar metallicity.  Because (sub-)mm continuum
observations of cold dust ($\sim$10\,K) in the outer disk ($\ge$50\,AU)
correlate very well with observations of inner dust disks
\citep{Andrews2005}, the entire disk, from 0.1 to 100 AU, is thought to
be dispersed almost simultaneously (within $\sim$$10^5$\,yr).
Therefore, the lifetime of the entire disk is also thought to be
5--10\,Myrs.

The disk lifetime in a low-metallicity environment is of special
interest because of the discovery that the probability of a star hosting
a planet increases with stellar metallicity (''planet-metallicity
correlation'') \citep{Gonzalez1997,Fischer2005,Santos2003}.  However,
the disk lifetime in an environment with lower metallicity than the
solar neighborhood has not been studied.  By directly studying the disk
lifetime in the EOG, it may be possible to obtain clues to the mechanism
of this ``planet-metallicity correlation.''  Although the disk fraction
has been studied for a low-metallicity ([O/H]$\simeq$$-0.5$)
\citep{Arnault1988} star-forming cluster, 30 Doradus, in the Large
Magellanic Cloud (LMC) \citep{DF_LMC1,DF_LMC2}, the target stars were
limited to only very massive ones ($\sim$20$\,M_\odot$) because of the
limited sensitivity and spatial resolution for this extragalactic
cluster.  Because the EOG is much closer than nearby dwarf galaxies, it
allows us to obtain accurate disk fractions in a low-metallicity
environment to $\sim$0.3$\,M_\odot$, which is below the characteristic
peak of IMF \citep{Elmegreen2008ApJ}. This mass detection limit
corresponds to the lowest mass of the target stars for the
planet-metallicity correlation (e.g., FGK-type stars)
\citep{Fischer2005} and is therefore critical for understanding the
correlation.

\section{OBSERVATION} \label{sec:observation}

Our targets, the Digel Cloud 2 clusters \citep{My2008ApJ,NK2008} are
among the most distant embedded clusters in the EOG at $R_g \sim
19$\,kpc. They are located in a giant molecular cloud, Cloud 2, one of
the most distant molecular clouds in the EOG \citep{Digel1994}.
The metallicity [O/H] of Cloud 2 is measured at $-0.7$\,dex from the
radio molecular emission lines \citep{Lubowich2004}, which is comparable
to that of LMC ($\sim$$-0.5$\,dex) and Small Magellanic Cloud (SMC;
$\sim$$-$0.9\,dex) \citep{Arnault1988}.  The dust-to-gas ratio in Cloud
2 was estimated to be as low as 0.001 from a submm continuum observation
\citep{Ruffle2007}.  This is about 10 times smaller than in the solar
neighborhood, and is consistent with the metallicity.  Because the Cloud
2 clusters are much closer ($D \simeq 12$\,kpc) than LMC/SMC ($D \sim
50$\,kpc), they are quite suitable for the study of the disk fraction in
a low-metallicity environment.  Their youth (most likely age of
$\sim$0.5\,Myr)\color{black} \citep{My2006ApJ,My2008ApJ,NK2008} adds an
advantage to the present study because any variation in disk fraction
should be most clearly seen when the disk fraction value is near its
maximum.  Because the EOG is located at the large heliocentric distance
of $\sim$10\,kpc, high sensitivity and high spatial resolution is
necessary for the present study.

Deep $JHK_S$-band images of Cloud 2 clusters were obtained on September
2, 2006 with an exposure time of $\sim $600\,sec for each band with the
8.2-m Subaru telescope equipped with a wide-field NIR camera, MOIRCS
\citep{MOIRCS}.  The seeing was as good as $\sim$0.35$''$--0.45$''$
through the night.  As a result, we obtained deeper images than our
preliminary studies of the Cloud 2 clusters \citep{My2006ApJ,My2008ApJ}.
All the data were reduced with IRAF\footnote{IRAF is distributed by the
National Optical Astronomy Observatories, which are operated by the
Association of Universities for Research in Astronomy, Inc., under
cooperative agreement with the National Science Foundation.}  with
standard procedures, and the results are shown in Fig.~\ref{fig:3colNS}.
$JHK$ aperture photometry has been performed using IRAF apphot.  The
Mauna Kea Observatories (MKO) filters were used \citep{Tokunaga2002}.
GSPC P330-E, which is a MKO standard \citep{Leggett2006}, was used for
the photometric calibration.  Because the fields of the Cloud 2 clusters
are very crowded, we used the aperture diameters $\sim$$0.6''$ and
$0.5''$ for the Cloud 2-N and 2-S cluster members, respectively, to
avoid the contamination of adjacent stars.  The limiting magnitudes
(10\,$\sigma$) for the Cloud 2-N cluster are $J=22.3$\,mag,
$H=21.7$\,mag, $K_S=21.2$\,mag, while those for the Cloud 2-S clusters
are $J=22.2$\,mag, $H=21.3$\,mag, $K=21.0$\,mag.  The achieved limiting
magnitudes correspond to the mass detection limit of
$\sim$0.1\,$M_\odot$, which is similar to the mass detection limit of
extensive observations of embedded clusters in the solar neighborhood
with smaller (2--4\,m class) telescopes.  Therefore, we can compare the
disk fractions in the low-metallicity environment to those in the solar
neighborhood on the same basis for the first time.

\section{IDENTIFICATION OF CLUSTER MEMBERS} \label{sec:identification}

\begin{figure}[htbp]
 \begin{center}
  \includegraphics[scale=0.83] {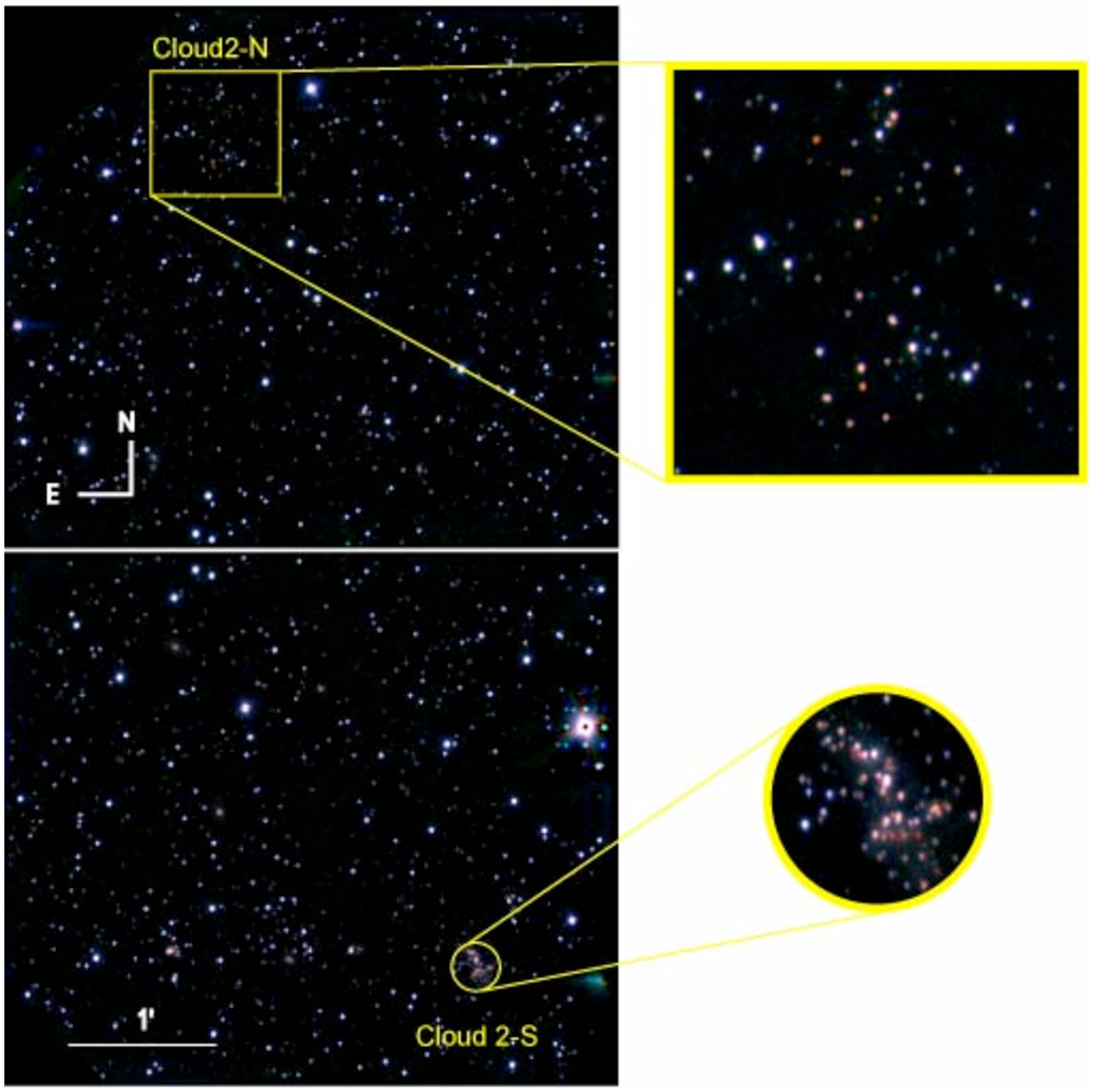}
\vspace{-45mm}
\caption{NIR pseudocolor images of the Cloud 2 clusters at the Galactic
radius of $R_g$$\sim$19\,kpc.  The color images of Cloud 2-N ({\it top})
and 2-S clusters ({\it bottom}) are produced by combining the $J$-
(1.26\,$\mu$m), $H$- (1.64\,$\mu$m), and $K_S$-band (2.14\,$\mu$m)
images obtained at the Subaru telescope on September 2008.  The field of
view of both images is $\sim$3.5$'\times$4$'$.  The yellow box and
circle mark the locations of the clusters, with closeups shown on the
right.}  \label{fig:3colNS}
 \end{center}
\end{figure}

\begin{figure}[htbp]
 \begin{center}
  \epsscale{1.0} 
  \plottwo{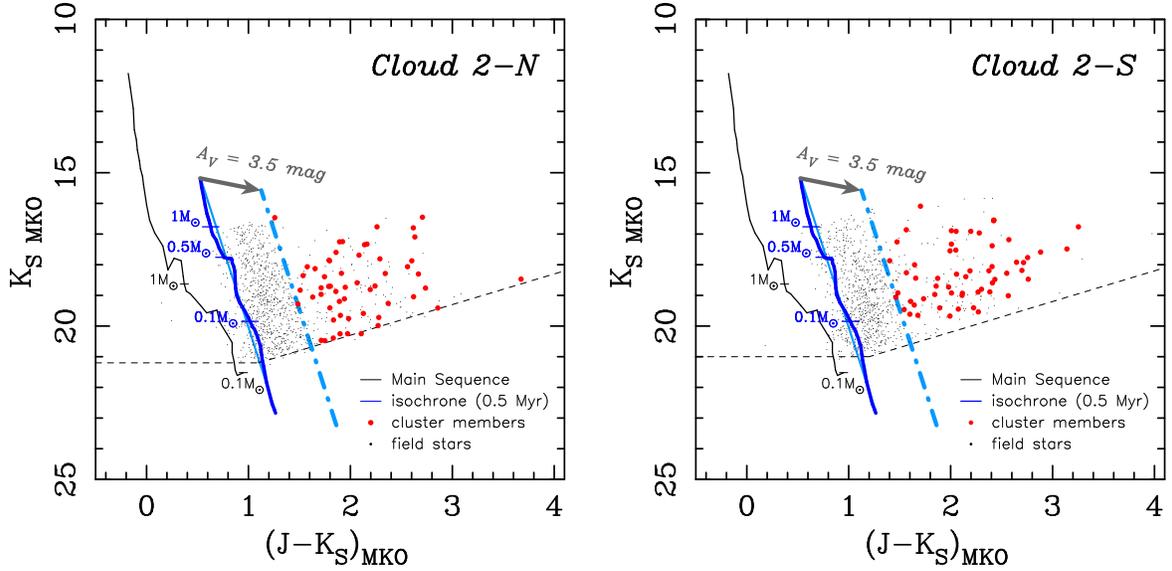}{f2b.eps}
\caption{ ($J-K_S$) vs. $K_S$ color-magnitude diagrams (CMD) of the
Cloud 2-N cluster ({\it left}) and the Cloud 2-S cluster ({\it right}).
Identified cluster members in the cluster regions (yellow box and circle
in Fig.~\ref{fig:3colNS}) are shown with red circles while all the
sources outside the cluster regions ($\sim$3.5$'\times$4$'$ field for
each cluster) are shown with black dots.  Only stars detected with more
than 10\,$\sigma$ in all $JHK_S$ bands are plotted.  The black lines
show the dwarf tracks \citep{Bessell1988}, while the blue lines show
isochrone models \citep[0.5\,Myr,][]{{D'Antona1997},{D'Antona1998}}.
For convenience, the isochrone model was approximated by the straight
line shown with a solid aqua line.  The gray arrows show the reddening
vectors of $A_V=3.5$\,mag from the isochrone model, while the dot-dashed
aqua lines show the isochrone model with the extinction.  Stars that are
in the Cloud 2 clusters region on the sky and are located to the right
of the dot-dashed lines on the CMD, are identified as the cluster
members.  The dashed lines show the limiting magnitudes (10\,$\sigma$).}
\label{fig:col-mag}
 \end{center}
\end{figure}

Because the Cloud 2 clusters are located behind the Cloud 2 molecular
cloud \citep{My2006ApJ}, the cluster members can be clearly identified
as extinct red sources behind the CO dense cores on the sky (see
Fig.~\ref{fig:3colNS}; see also Fig.~1 in Kobayashi et al. 2008). 
We identified cluster members of the Cloud 2-N and 2-S clusters with the
following criteria: (1) distributed in regions of the Cloud 2 clusters,
which are defined by the yellow box and circle in Fig.~\ref{fig:3colNS},
and (2) large $A_V$ excess compared to normal field stars (extinction of
$A_V\geq3.5$; see Fig. 2). On the color-magnitude diagram they are
located in the region apart from the region of normal field stars
because the Cloud 2 clusters are located at the peak of the Cloud 2
molecular cloud core (Kobayashi et al. 2008; Yasui et al. 2008) and
reddened by the large extinction. We confirmed that stars with
$A_V\geq3.5$\,mag are concentrated on the cluster fields, while stars
with $A_V< 3.5$\,mag are widely distributed over the observed field. In
Fig.~\ref{fig:col-mag} the identified cluster members in the cluster
regions are shown with red circles while all the sources outside the
cluster regions ($\sim$3.5$'\times$4$'$ region around each cluster) are
shown with black dots.  As a result, we identified 52 and 59 cluster
members for the Cloud 2-N and 2-S clusters, respectively, with the
10\,$\sigma$ limiting magnitude (red circles in Fig.~2) in both $K_S$
and $H$ bands. 

\begin{figure}[htbp]
\begin{center}
\epsscale{1.0} 
 \plottwo{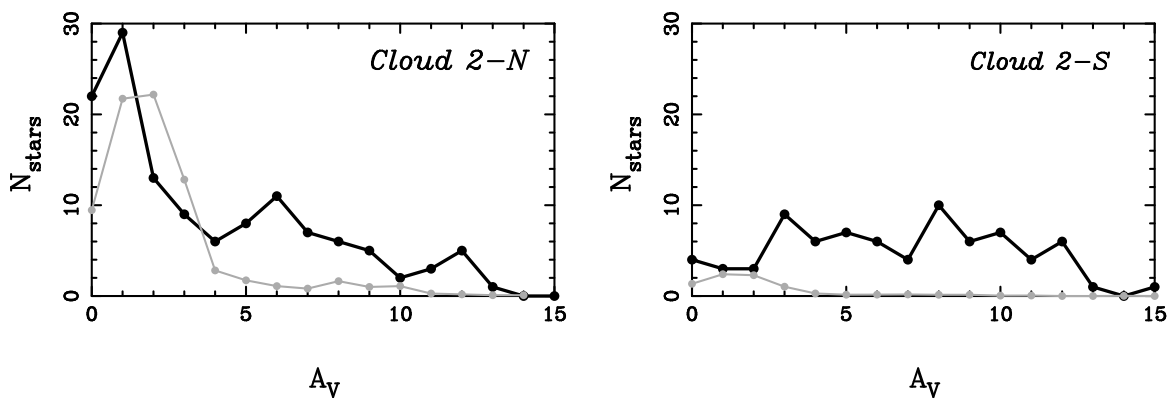}{f3b.eps}
\caption{ $A_V$ distributions of all stars in the cluster region
(cluster members; black dots and line) and all stars outside the cluster
region (field objects; gray dots and line). The $A_V$ is estimated from
Fig~\ref{fig:col-mag} based on $J-K$ colors.  The distribution of field
objects is normalized to match with the total area of the cluster
regions. Because the number of field objects decreases significantly at
$A_V\gtrsim 3.5$ mag, most cluster members can be distinguished from the
field objects as red sources with $A_V \geq 3.5$\,mag.}
\label{fig:av}
\end{center}
\end{figure}

Considering the very large $R_g$ of the Cloud 2 clusters
($R_g\simeq19$\,kpc), contamination of background stars is negligible
and most of the field objects are foreground stars (some are background
galaxies).  To quantify the contamination by foreground stars, we
compared the $A_V$ distributions of all the sources in the cluster
regions and the field objects outside the cluster regions
(Fig.~\ref{fig:av}).  Because the number of field objects decreases
significantly at $A_V\sim3.5$\,mag, most cluster members can be
distinguished from the field objects as red sources with $A_V \geq
3.5$\,mag. The contamination by the foreground stars are estimated by
counting the field objects in the tail of the distribution at $A_V \geq
3.5$\,mag and divide it with the total number of the sources in the
cluster regions: $\sim$20\,\% and $\sim$2\,\% for the Cloud 2-N and 2-S
clusters, respectively. Note that the contamination is very small for
the S cluster because of the smaller area of the cluster
region. Although there must be some cluster members at $A_V<3.5$ that
missed our identification, the number of such sources should not be
significant in view of the geometry of the molecular clouds, which are
in front of the clusters (Yasui et al. 2006, 2008).

\section{DISK FRACTION}
\label{sec:diskfraction}

\begin{figure}
\begin{center}
 \includegraphics[scale=0.50]{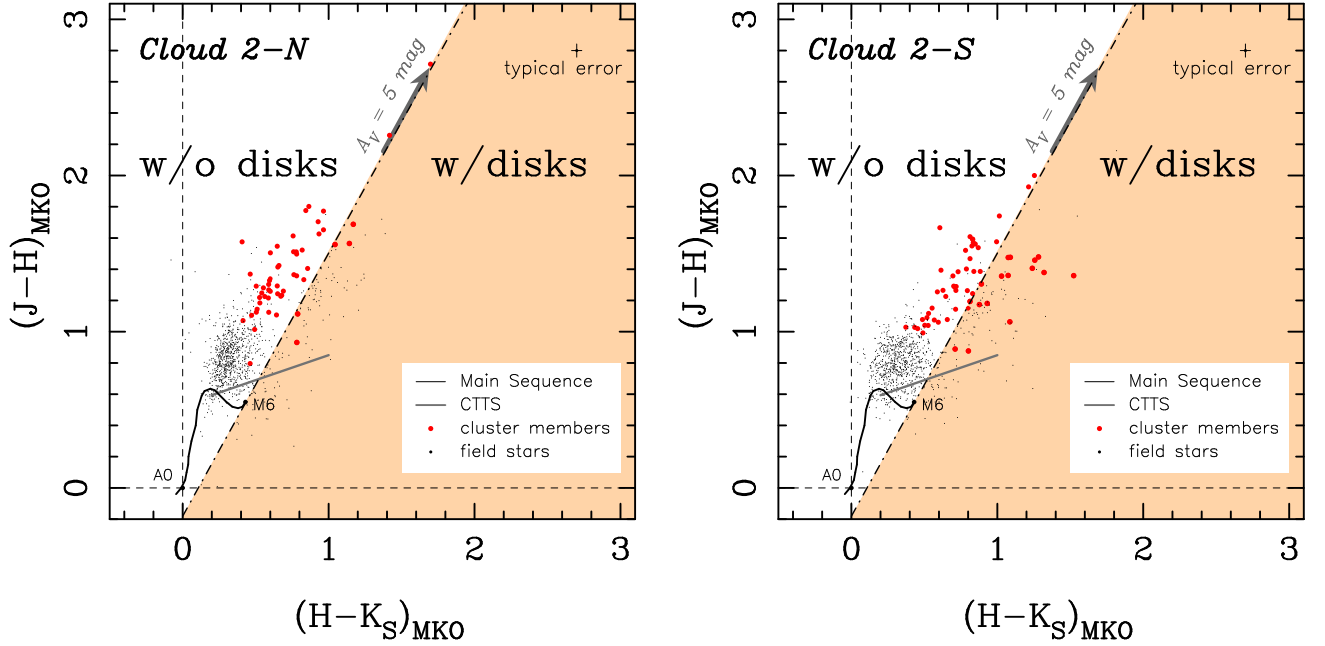}
\caption{ ($J$$-$$H$) vs. $(H$$-$$K)$ color-color diagrams for the Cloud
2-N cluster ({\it left}) and the Cloud 2-S cluster ({\it right}) in the
MKO system.  Identified cluster members and field stars are shown with
red filled circles and black dots, respectively. Typical uncertainty
(1\,$\sigma$) of the colors are shown at the top-right corner.  The
solid curve in the lower left portion of each diagram is the locus of
points corresponding to the unreddened main-sequence stars.  The
dot-dashed line, which intersects the main-sequence curve at the maximum
{\it H}$-$$K_S$ values (M6 point on the curve) and is parallel to the
reddening vector, is the border between stars with and without
circumstellar disks.  The classical T Tauri star (CTTS) loci are shown
with gray lines.}

\label{fig:colcolNS}
\end{center}
\end{figure}

We derived disk fractions of the Cloud 2 clusters using a {\it J-H}
vs. {\it H-K} color-color diagram for the identified cluster members
(Fig.~\ref{fig:colcolNS}).  Stars with circumstellar dust disks are
known to show a large $H-K$ color excess \citep{Lada1992}.  On the $JHK$
color-color diagram, stars without circumstellar disks are seen as
main-sequence stars reddened with extinction, while stars with
circumstellar disks are seen in ``the disk excess region'' (to the right
of the dot-dashed line in (Fig.~\ref{fig:colcolNS}) because of thermal
emission from hot dust disk with a temperature of 1000--1500\,K.  For
the Cloud 2-N and 2-S clusters, the disk fractions are 9$\pm$4 \% $(5 /
52)$ and 27$\pm$7 \% $(16 / 59)$, respectively, where the ratio of stars
with excess to the total number of stars is given in parentheses.  The
uncertainty of the disk fractions is estimated only from the Poisson
errors because photometric errors are quite small with the median errors
of 0.06\,mag for ($H-K$) and ($J-H$) \citep{Liu2003}.  Given the
possible systematic uncertainty, we cannot say at this stage if the
difference between the disk fractions of the Cloud 2-N and 2-S is
significant.
Because the above-discussed contamination from the foreground objects of
$A_V \geq 3.5$ mag may decrease the disk fraction, the actual disk
fractions could be higher than the derived values here. However, the
amount of the increase is estimated as $+$2.2\,\% and $+$0.5\,\% for the
Cloud 2-N and -S clusters, respectively, when reducing the total number
of the identified cluster members (52 and 59 for the N and S clusters,
respectively) by the estimated number of contaminated foreground objects
($\sim$10 and $\sim$1 for the N and S clusters, respectively). Because
this maximum possible increase of the disk fraction is well within the
large Poisson uncertainty (4\,\% and 7\,\% for the Cloud 2-N and 2-S
clusters, respectively), we conclude that the effect of the background
contamination on the disk fraction is small. In addition, in the case
where we assume that all stars in the cluster regions are cluster
members, the disk fractions for the Cloud 2-N and 2-S cluster become
6\,\% (5/89) and 21\,\% (16/78), respectively.
Even in this extreme case, the amounts of the decrease ($-3$\,\% for the
Cloud 2-N cluster and $-6$\,\% for the 2-S cluster) are again within the
Poisson errors.

To confirm the negligible effect of the photometric errors, we estimated
disk fractions of two extreme cases: one case is that ($J-H$) and
($H-K$) colors of all stars were systematically changed within the
photometric uncertainties of each source to get the lowest disk fraction
and the other case to get the highest disk fraction.
Assuming a Gaussian distribution of the photometric errors, the
probability that the color of a star is within the $\pm$1\,$\sigma$
offset is 68\,\% and the probability that the color of a star is redder
than the 1\,$\sigma$ offset is about 16\,\% ($=(100 - 68)/2$).  Because
the Cloud 2 clusters have about 50 members, the probability that {\it
all} stars have redder colors than the 1$\sigma$ offset is $~0.16^{50}$,
which is essentially zero. If we assume that the probability
distribution of disk fraction is Gaussian, the estimated disk fraction
of the extreme case (probability 0.16$^{50}$) should have a probability
less than the 3 sigma case (0.1\,\%).
In the above extreme case (all the sources have $+1\sigma$ or $-1\sigma$
redder colors), we found that the disk fraction increases from 9\,\% to
19\,\% (N cluster) and from 27\,\% to 44\,\% (S cluster).  
Since such an extreme case only occurs with a probability of
$0.16^{50}$, much less than a 3 sigma occurrence, we consider that the
1\,$\sigma$ photometric uncertainty must be much less than 3\,\%
($(19-9)/3 \sim 3$) and 6\,\% ($(44-27)/3 \sim 6$) for the Cloud 2-N and
2-S clusters, respectively, which is smaller than the Poisson errors.

In view of the low-metallicity environment of the EOG, there might be
some fundamental differences between the clusters in the solar
neighborhood and in the EOG, such as the reddening and the stellar mass
distributions, that may require different treatments of the disk
fraction estimation. In our earlier paper, we confirmed that the
reddening vector as well as the main-sequence track on the color-color
diagram does not change with metallicity down to the metallicity of the
Cloud 2 (Yasui et al. 2008, section~4). Also, the stellar mass
distributions of the Cloud clusters are suggested to follow the
universal IMF, which is widely observed in the solar neighborhood (Yasui
et al. 2006, 2008). Therefore, we conclude that the use of the same
division of with/without disk as for the solar neighborhood is
appropriate.  As for the treatment of the different filter systems, see
Yasui et al. (2008) for the detail.

\section{$JHK$ DISK FRACTION OF THE NEARBY EMBEDDED CLUSTERS}
\label{sec:JHK_DF}

We derived the $JHK$ disk fraction of various embedded clusters in the
solar neighborhood using the photometry data in the literature (Table~1
and Fig.~\ref{fig:DFcl2}).  We chose publications with the following
criteria:
i) $JHK$ photometry data of cluster members are available in order to
estimate disk fractions in the same way as for our data,
ii) the mass detection limit is $\le$1\,$M_\odot$ \citep{Haisch2001ApJL},
iii) all data in each embedded cluster are estimated in the same filter
system.
For all clusters we used the classical dwarf track in the Johnson-Glass
system \citep{Bessell1988} and a reddening vector in the Arizona system
\citep {RL}.  $JHK$ disk fractions were estimated in the same way as for
the Cloud 2 clusters: the border line, which determines whether a star
has a disk or doesn't have a disk, passes through the point of M6 type
stars in the dwarf track on the color-color diagram.

Recent disk fraction studies often use $JHKL$ data because thermal
infrared ($\lambda \geq 3$\,$\mu$m) can trace the colder (thus, outer)
disk, and disk fractions from $JHKL$ data are thought to be a more
robust signature of the circumstellar disk than those only from $JHK$
data \citep{Haisch2001ApJL}.  However, disk fractions from only $JHK$
data are about 0.6 of those from $JHKL$ data with slightly more scatter
on the disk fraction-age plot, and the estimated disk lifetime from
$JHK$ data is basically identical to that from $JHKL$ data.  Therefore,
despite a little larger uncertainty, the disk fraction from only $JHK$
data should still be effective even without thermal IR data.  Since the
detection sensitivity in the thermal infrared is still not enough for
the distant clusters like those in the EOG, we can make use of the high
detection sensitivity at $JHK$ wavelengths to estimate the disk lifetime
for the EOG clusters.


\section{DISCUSSION}
\color{black}

\subsection{Disk Fraction in Low-metallicity Environment}

We show the disk fractions of the Cloud 2 clusters, as well as of the
derived disk fraction of seven embedded clusters in the solar
neighborhood, in the disk fraction-age diagram (Fig.~\ref{fig:DFcl2};
see $\S$~\ref{sec:JHK_DF}).
The disk fractions of the Cloud 2-N and 2-S clusters are significantly
lower than nearby embedded clusters of similar age.  As discussed in
Section \ref{sec:diskfraction}, the contamination from the foreground
objects and the possible unidentified cluster members have only a small
effect on the estimated disk fraction.
Although the difference between the filter system for our data (MKO) and
the data in the literature might cause some systematic offset for the
disk fraction value, we confirmed that the disk fraction does not depend
on the choice of the filter system by confirming that the disk fraction
of a good reference cluster (NGC 2024 cluster) is identical for both CIT
and MKO systems (see Appendix~A).

Given the importance of the age of the clusters to the above conclusion,
here we discuss the uncertainties of the age and the age spread of the
clusters. Considering the IMF derived from $K$-band luminosity function
(KLF), the age of the Cloud 2 clusters should be less than 2\,Myr: the
KLF fitting shows an unrealistic top-heavy IMF with the age of 2\,Myr
\citep{My2006ApJ, My2008ApJ}.  Therefore, we conservatively estimate
that the age upper-limit of the Cloud 2 clusters is 1\,Myr.  Moreover,
the suggested supernovae triggered star formation of the Cloud 2
clusters \citep{NK2008} suggests the age of 0.4\,Myr, which strongly
supports the above estimate from the KLF. Because there is no
spectroscopic study of the cluster members, there is no observed age
spread information as for the nearby clusters as in e.g.,
\citet{Palla2000}.  Because the mean age of this cluster is estimated as
$\sim$0.5\,Myr, the age spread should be smaller down to $\pm$0.5\,Myr
as assumed in the young cluster model in Muench et al. (2000). Muench et
al. (2002) estimated the age of the Trapezium cluster at
0.8$\pm$0.6\,Myr.  Palla \& Stahler (2000) actually obtained smaller age
spread for the youngest embedded clusters (rho Oph, ONC). Moreover, in
the above triggered star formation picture (Kobayashi et al. 2008), the
age spread of the Cloud 2 clusters should be small because of the single
triggering mechanism at a certain time.  Therefore, we conclude that the
most likely age of the clusters is 0.5\,Myr and no more than 1\,Myr.
Even if we consider the possible age range, the estimated disk fractions
are still quite low compared with those in the solar neighborhood.


Although the Cloud 2 clusters appear to be quite close on the sky, they
are totally independent clusters with a projected distance of
$\sim$25\,pc at the heliocentric distance of $D \simeq12$\,kpc
\citep{My2008ApJ}.  If they were located at the same heliocentric
distance as the Orion dark cloud ($D \simeq 400$\,pc), the angular
separation between the clusters becomes $3.5^\circ$, which is larger
than $2.5^\circ$ between two independent clusters, e.g., NGC 2024 and
NGC 2071 clusters, listed in the previous study of $JHK$ disk fraction
\citep{Lada1999}.  Because the measured disk fractions for two {\it
independent} clusters were similarly low, the disk fractions of embedded
clusters in the low-metallicity environment are inferred to be
universally low, at least in the EOG.  In combination with the fact that
{\it there are no very young ($\leq$1\,Myr) embedded clusters of solar
metallicity whose disk fractions are this low}, our results strongly
suggest that disk fraction depends strongly on metallicity.

\subsection{Why Lower Disk Fraction with Lower Metallicity?}
\label{sec:why_lowDF}

The most simple interpretations of the low disk fraction are that the
disk is optically thin either because of the collisional agglomeration
of dust grains in the disk \citep{Andrews2005} or because of the low
dust-to-gas ratio due to the low metallicity.  However, the former is
unlikely because the probability of dust collision, which is
proportional to the square of dust density, should be rather low in the
low-metallicity environment.  The latter is also unlikely because the
inner disk should be optically thick even with a metallicity of
$-1$\,dex: opacity of the disk with the wavelength of 2\,$\mu$m ($\nu
\simeq 1.5 \times 10^{14}$\,[Hz]) at 0.1\,AU is estimated to be
$\sim$10$^5$, assuming a minimum mass solar nebula (surface gas density
of $\simeq$5$\times$10$^4$\,g$/$cm$^2$ at 0.1\,AU) \citep{Hayashi1981}
and a standard power-law opacity coefficient $\kappa_\nu$ ($\simeq$15;
$\kappa_\nu = 0.1 (\nu / 10^{12}\,{\rm [Hz]})$\,cm$^2$\,g$^{-1}$)
\citep{Beckwith1990}, and considering the dust-to-gas ratio of 0.001
\citep{Ruffle2007}, which is 1/10 smaller than in the solar metallicity.
If we assume that the typical disk-to-star mass ratio (0.01; e.g.,
Andrews \& Williams 2005, Natta et al. 2006), we would expect the total
disk mass of $10^{-3}$\,$M_\odot$ for the lowest-mass stars in our
sample ($M_{*}\sim0.1$\,$M_\odot$). Because this is roughly one-tenth
that of the minimum mass solar nebulae (0.01--0.02\,$M_\odot$), we still
expect a very large optical thickness of $\sim$$10^4$ at K-band,
assuming that disk mass measured in the submm correlates directly with
surface density in the inner regions.  This assumption is used for disks
in the solar metallicity environment \citep[e.g., ][]{Andrews2005}.
Even if we assume the lowest observed disk-to-star mass ratio in the
nearby star forming regions ($\sim$$10^{-4}$; Andrews \& Williams 2005),
the expected optical thickness is still $\sim$$10^2$.  Model spectral
energy distributions with various disk mass do not show a significantly
different $H-K$ color excess (less than 0.05 mag) even if the amount of
dust disks decreased to $1/10$ \citep{Wood2002}, which correspond to the
dust-to-gas ratio of $\sim$0.001 in the EOG.  Therefore, we conclude
that the small disk fraction implies that the inner region of the disk
is cleared out.

Note that optically thickness does not necessarily mean the existence of
observable $H-K$ excess for lowest-mass stars with low effective
temperature ($\sim$3000\,K) and with very low disk mass of
$10^{-3}$--$10^{-8}$\,$M_\odot$ \citep[e.g.,][]{Ercolano2009}.  Because
the lowest mass stars of our samples ($\sim$0.1\,$M_\odot$) are M6-type
stars with an effective temperature ($T_{\rm eff}$) of $\sim$3000\,K
\citep[see e.g., Fig.~7 bottom figure in][]{LadaLada2003}, it may be
inevitable that some lowest-mass stars in our sample do not have
detectable $H-K$ excess simply because of the less dust content under
low-metallicity, thus contributing to the low disk fraction
artificially.
To investigate this effect, we made a plot of the disk fraction of the
Cloud 2 clusters for all sources brighter than certain $K$-band
magnitudes (Fig.~\ref{fig:DFmag}). It is clearly seen that the disk
fractions stay almost flat at the low level and do not {\it decrease}
even including the M stars of the faintest magnitudes ($K\sim20$) that
may lower the disk fraction as described above. Therefore, we conclude
that the disk fraction is genuinely low for all sources down to the mass
detection limit and that clearing out of the inner region of the disk is
the reason for it.  For the above arguments, we note two related
studies. One is the detailed disk fraction study of the Trapezium
cluster by \citet{Lada2004AJ}, who show that even for the latest M-type
stars (e.g., M9-type stars, $\sim$0.02\,$M_\odot$, $T_{\rm
eff}\sim2600$\,K at 1\,Myr) has detectable $H-K$ excess and the $JHK$
disk fraction does not change even including the stars of this low-mass.
In view of the youth of the Cloud 2 clusters ($\sim$0.5\,Myr) that is
similar to the Trapezium cluster, we would naively expect similar
detectable K-band excess even for the lowest mass stars in our sample
($\sim$0.1\,$M_\odot$, M6-type stars) with $T_{\rm eff}\sim3000$\,K even
with the low metallicity. Also, in the simulation by \cite{Wood2002}, a
detectable $H-K$ excess is seen even with very small disk mass of
$10^{-6}$\,$M_\odot$ under solar metallicity (which corresponds to
$10^{-5}$\,$M_\odot$ with the metallicity of Cloud2), assuming a central
source with 0.6\,$M_\odot$ and $T_{\rm eff}=4000$\,K.

Why does the inner region of the disk becomes cleared out effectively
with lower metallicity?  There are two well-known mechanisms that can
clear the inner disk: dust sublimation \citep{Millan-Gabet2007} and the
X-wind model \citep{Shu1994}.  In the following, we discuss if these
mechanisms can effectively work with lower metallicity.
 {\it i) Dust sublimation:} Since dust sublimation typically occurs
around 1500\,K, the very inner part of the disk with $T >$1500\,K
becomes dust-free.  The dust sublimation radius, $R_{\rm s}$, is
described approximately as $( L_* / 4 \pi T_{\rm rim}^4 \sigma)^{1/2}$,
where $L_*$ is stellar luminosity, $T_{\rm rim}$ ($\sim$1500\,K) is disk
temperature at inner disk radius ($R_{\rm s}$), and $\sigma$ is the
Stefan-Boltzmann constant \citep{Dullemond2007}.  However, because the
inner disk is expected to be optically thick even with the low
metallicity of $-1$\,dex, the temperature distribution inside the disk
for $-1$\,dex should not differ from that for the solar metallicity
(0\,dex).  Therefore, $R_{\rm s}$ is not expected to be significantly
larger in the lower-metallicity environment.
{\it ii) X-wind model:} Due to the interaction between the stellar
magnetic field with the accretion disk, the gas disk is thought to be
truncated at an inner radius of $R_{\rm x} = \Phi_{\rm dx}^{-4/7}
(\mu_*^4 / G M_* \dot{M_D^2})^{1/7}$ \citep{Shu2000}, where $M_*$ is the
star's mass, $\dot{M_D}$ is the disk accretion rate, $\mu_*$ is magnetic
dipole moment, $\Phi_{\rm dx}$ is a dimensionless number of order unity,
and $G$ is the gravitational constant.  Among these parameters
determining $R_{\rm x}$, only the disk accretion rate, $\dot{M_D}$
depends on metallicity. The magnetorotational instability (MRI), which
is the most likely cause for disk accretion \citep{Hartmann2009book},
largely depends on the ionization of the gas disk by Galactic cosmic
rays \citep{Gammie1996ApJ}, X-rays from the central young star
\citep{Igea1999}, and thermal ionization \citep{Sano2000}.  In a
low-metallicity environment, we can expect that the ionization fraction
increases and that the dead zone, which is the MRI inactive region of
the disk (Gammie 1996), shrinks because the recombination on grain
surfaces is reduced. As a result, $\dot{M_D}$ increases and $R_{\rm x}$
decreases.  In fact, \citet{Sano2000} suggested that the dead zone
shrinks in the case that the dust sedimentation of dust grain proceeds,
which mimics the case in the low-metallicity.  On the other hand,
low-metallicity also means fewer molecular ions, which would reduce the
ionization fraction \citep[e.g.,][]{Sano2000}, and thus would reduce
$\dot{M_D}$.  Which process dominates in the low metallicity environment
has not been determined conclusively since the combination of many
physical processes determine the size of the dead zone (e.g, Fromang,
Terquem, \& Balbus 2002).  However, the former process, thus larger
$\dot{M_D}$, may dominate (T. Sano, T. Suzuki, F. Shu, private
communication).

Although further theoretical studies are necessary to have a firm
conclusion, we would naively expect that $R_{\rm x}$ is not
significantly larger in a lower-metallicity environment, and it appears
that there is no theoretical reason to suspect that lower metallicity
would make disks less detectable through the H-K excess. The lower dust
level would still result in an optically thick inner edge and the size
of the cleared out inner region should not change significantly with the
metallicity.  Therefore, our observations can simply be interpreted that
the inner disk (gas and dust) is mostly absent in Cloud 2-N and 2-S.

\subsection{Short Disk Lifetime in a Low-metallicity Environment}

\begin{figure}[htbp]
 \begin{center}
  \includegraphics[scale=0.60] {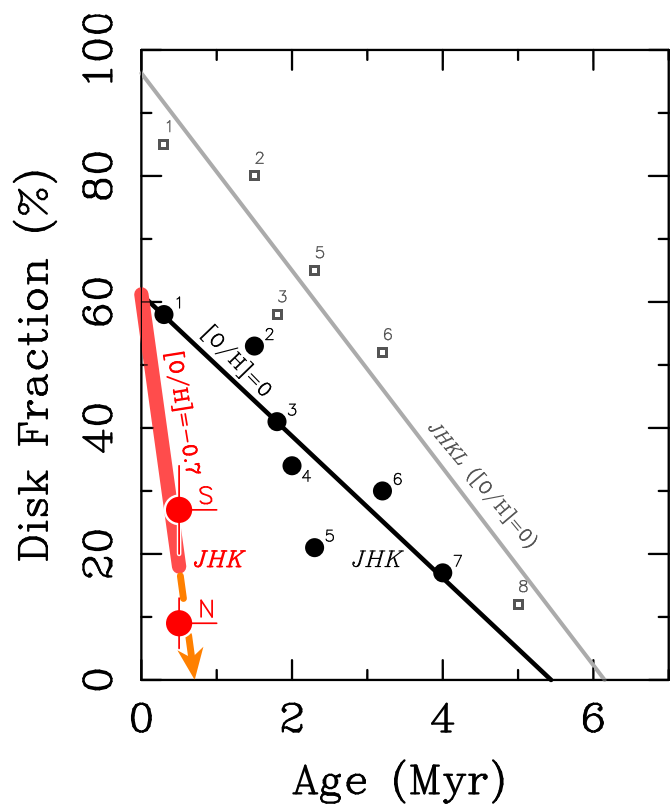}
\caption{Disk fraction as a function of cluster age.  $JHK$ disk
fractions of the Cloud 2 clusters are shown with red filled circles,
while derived $JHK$ disk fractions of other young embedded clusters with
solar metallicity are shown with black filled circles (see
$\S$~\ref{sec:JHK_DF}).  The $JHKL$ disk fractions of those other
clusters \citep{Haisch2001ApJL} are shown with gray open squares.  The
black and gray lines show the disk fraction evolution under the solar
metallicity, while the red line shows the proposed $JHK$ disk fraction
evolution in the low-metallicity environment.  These lines are derived
with least-squares fits.}  \label{fig:DFcl2}
\end{center}
\end{figure}

\begin{figure}[htbp]
 \begin{center}
  \includegraphics[scale=0.50] {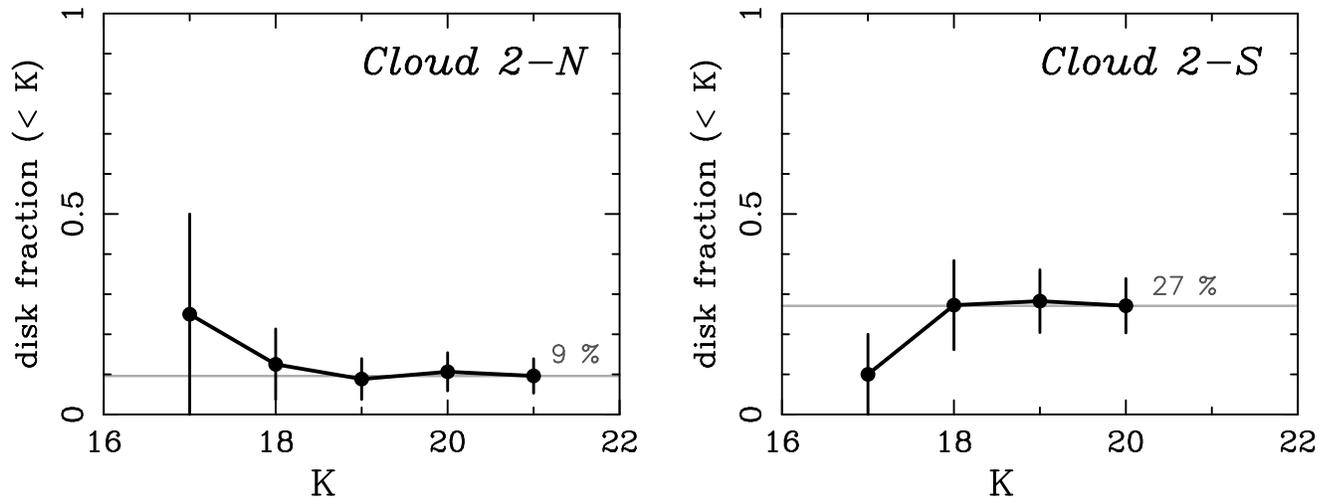}
\caption{Disk fraction as a function of the $K$-band manitude. Disk
fractions are derived for all the sources brighter than the designated
magnitude. It is clearly seen that the disk fractions stay almost flat
at the low level and do not decrease within the uncertainty even
including the late M stars at the faintest magnitudes ($K\sim20$).  }
\label{fig:DFmag}
\end{center}
\end{figure}

 The above arguments lead to the suggestion that {\it the initial disk
fraction (age $=$ 0 in Fig.~\ref{fig:DFcl2}) in the low-metallicity
environment of our targets is expected to be as high as that for the
solar metallicity case}, assuming that the distribution of initial disk
properties is exactly the same regardless of metallicity. Although this
assumption is not obvious and more detailed observations of the disks in
the extreme outer Galaxy are needed, it is not an unreasonable
assumption because the star formation process from gas to stars do not
seem to significantly change for the metallicity of the Cloud 2
clusters, for example, in view of the universal IMF throughout various
physical/chemical environments (e.g., Elmegreen et al. 2008, see Omukai
et al. 2000 for theoretical study).  Therefore, we suggest that the
significant decrease of the disk fraction in Fig.~\ref{fig:DFcl2} is a
result of the much shorter lifetime ($\sim$1\,Myr) of the inner disk in
the low-metallicity environment compared to that in the solar
metallicity environment, 5--6\,Myr from $JHK$ disk fraction
\citep[$\S$~\ref{sec:JHK_DF}, ][]{Lada1999}.  This can be expressed with
the rapid decrease of the disk fraction as shown with the red thick line
in Fig.~\ref{fig:DFcl2}.

\begin{figure}[htbp]
 \begin{center}
  \epsscale{0.5}
  \plotone{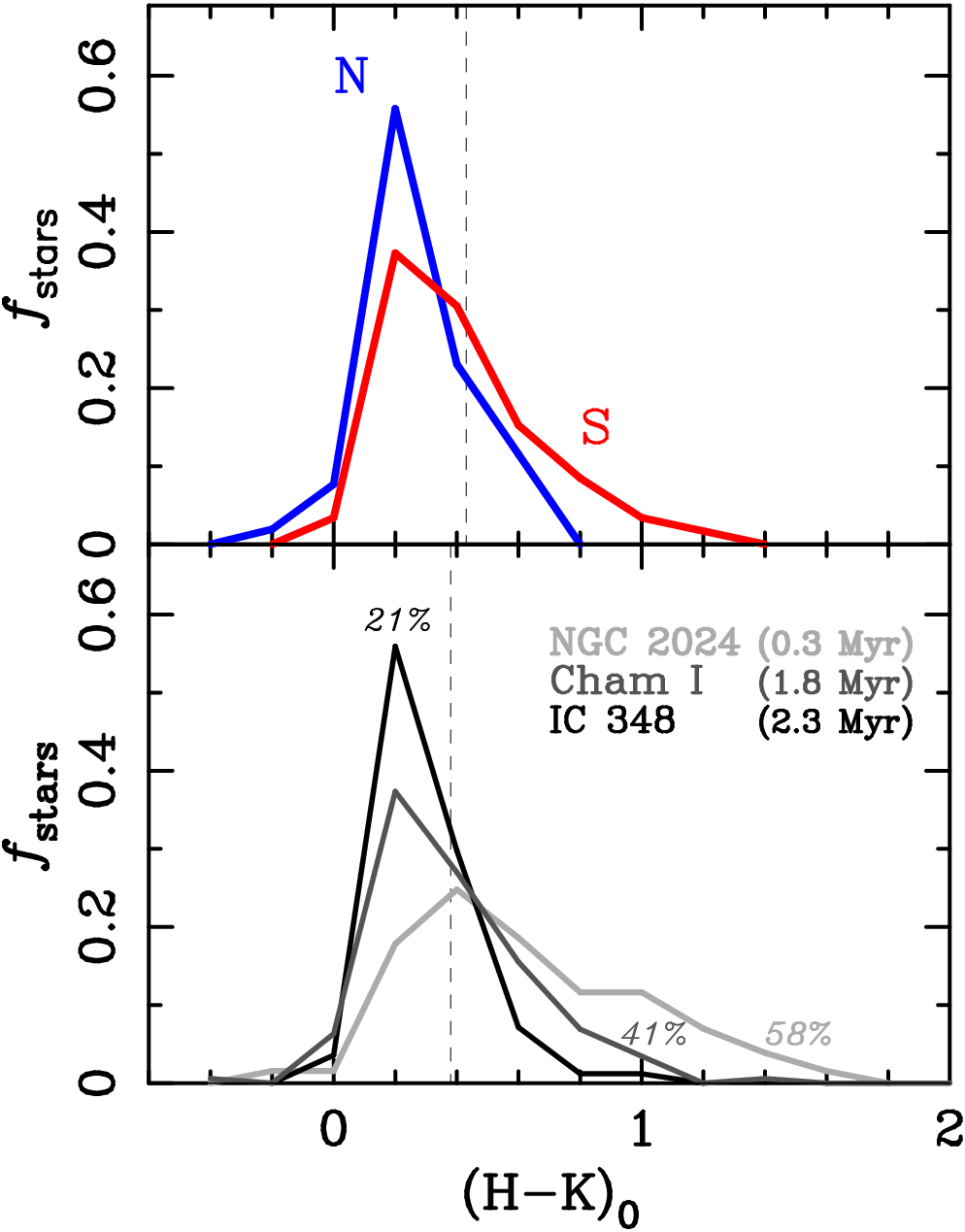}
\caption{ Comparison of intrinsic $H-K$ color distributions.  The
fraction of stars ($f_{\rm stars}$) per each intrinsic color bin
($(H-K)_0$) are plotted.  {\it Top}: $(H-K)_0$ distributions for the
Cloud 2 clusters in the MKO system.  {\it Bottom}: $(H-K)_0$
distributions for the various young clusters in the solar neighborhood:
NGC 2024 \citep{Haisch2000AJ}, Cham I \citep{Lhuman2004}, IC 348
\citep{Haisch2001AJ}.  Disk fractions of NGC 2024, Cham I, and IC 348
are 58\,\%, 41\,\%, and 21\,\%.  The dashed lines show the borderlines
for estimating the disk fraction.  Note that the border of the $(H-K)_0$
value for estimating the disk fraction (dashed line) differs slightly
for different filter system.}
\label{fig:HKdist}
 \end{center}
\end{figure}

We know that the entire circumstellar disk, both the inner and outer
disks, dissipates almost simultaneously in the solar metallicity
environment because the submillimeter detection fraction and the
fraction of objects with a NIR excess are identical \citep{Andrews2005},
and also because the observed fraction of transition objects, between
the Class II and Class III states, is very low \citep{Duvert2000}.
Therefore, we suggest that the {\it entire} disk dispersal is also rapid
in the low-metallicity environment.
The distributions of ({\it H}$-${\it K}) excess in the Cloud 2 clusters,
which are consistent with those of more evolved ($\sim$5\,Myr) embedded
clusters with solar metallicity, also support the above idea, though
other mechanisms might cause the formation of the similar $(H$$-$$K)$
distributions.
We constructed intrinsic $H-K$ color distributions
(Fig.~\ref{fig:HKdist}) for Cloud 2 clusters and nearby embedded 
clusters (Table~1).  The intrinsic $(H-K)$ colors of each star were
estimated by dereddening along the reddening vector to the young star
locus in the color-color diagram (see Fig.~\ref{fig:colcolNS}).  For
convenience the young star locus was approximated by the extension of
the CTTS locus (gray lines in Fig.~\ref{fig:colcolNS}), and only stars
that are above the CTTS locus were used.  For the Cloud 2 clusters
(Fig.~\ref{fig:HKdist}, top), the CTTS locus in the MKO system
\citep{My2008ApJ} was used, while for the nearby clusters
(Fig.~\ref{fig:HKdist}, bottom), the original CTTS locus in the CIT
system \citep{Meyer1997} was used.  To make the figure clearer, we show
the color distributions for only NGC 2024 \citep{Haisch2000AJ}, Taurus
\citep{Kenyon1995}, and IC 348 \citep{Haisch2001AJ}, which have
significantly different disk fractions.  It is clearly seen that the
distribution becomes bluer and sharper with lower disk fractions.  The
distributions of the Cloud 2 clusters resemble that of IC 348, whose
disk fraction is as low as the Cloud 2 clusters ($\sim$20\,\%).


\subsection{Physical Mechanism for the Rapid Disk Dispersal in a
Low-metallicity Environment}
\label{sec:mechanism}

As the mechanism of disk dispersal, two possibilities are known : (i)
disk accretion \citep{Harmann1998}, and (ii) photoevaporation of the
surface layer of the gas disk due to extreme-ultraviolet (EUV) radiation
from the central star and/or the external radiation field
\citep{Hollenbach2000PPIV}.
In the low metallicity environment, both possibilities may work
effectively.  First, the disk accretion rate may increase in the
low-metallicity as discussed in $\S$~\ref{sec:why_lowDF}, if the
increase of the ionization fraction due to less dust is larger than the
decrease of the ionization fraction due to less molecular ions.
This could be the reason for the rapid disk dispersal.
Secondly, the photoevaporation process may also work effectively to
shorten the disk lifetime.  Because the EUV photoevaporation is solely
based on the Str\"{o}mgren condition of the hydrogen gas, metallicity
(or dust-to-gas ratio) should not affect the rate of the
photoevaporation mass-loss.  However, the recent theoretical work
\citep{Gorti2009} shows that far-ultraviolet (FUV) or X-ray radiation is
more effective than EUV to photoevaporate the disk.  Because FUV
penetration strongly depends on the amount of dust in the disk, they
found that the disk lifetime decreases with the dust opacity.  Although
the dust opacity dependence of the disk lifetime is not strong (factor
of $\sim$2 changes with the dust opacity change of 10), they noted that
the results are still in a preliminary stage and has to wait further
detailed modeling.
Therefore, this could be also the reason for the rapid disk dispersal in
low-metallicity.  In addition, something specific to the EOG environment
might be the reason \citep[see e.g., ][]{Haywood2008}, but obviously
further studies are necessary to pin down the physics of this phenomena.


\subsection{Implication to the Planet-metallicity Correlation}
\label{sec:PMcor}

If the lifetimes of disks at the solar metallicity (${\rm [O/H]} \sim
0$) and at the low-metallicity (${\rm [O/H]} \sim -0.7$) are
approximated as 5\,Myr and 1\,Myr, respectively, our results suggest
that the disk lifetime strongly depends on the metallicity (Z $\equiv$
[M/H]) with a $\sim10^Z$ dependence.  Because the disk lifetime
($\tau_{\rm disk}$) is directly connected to the planet formation
probability (p$_{\rm pl}$), the strong metallicity dependence of
$\tau_{\rm disk}$ may create a similarly strong metallicity dependence
of p$_{\rm pl}$.  Therefore, the disk dispersal could be one of the
major driving mechanisms for the planet-metallicity correlation, which
shows a strong sharp ($\sim10^{2Z}$) metallicity dependence (Fischer et
al. 2005).
In the current standard core nucleated accretion model of the giant
planet formation \citep[e.g., see a review by][]{Lissauer_PPV}, the
planet formation time scale is determined both by the core growth time
and the gas accretion time onto the core. Since the first study by
Safronov (1969), core growth time in the giant planet zone has been
thought to be far longer than the observed $\tau_{\rm disk}$. But, after
the discovery that $\tau_{\rm disk}$ is less than 10\,Myr, it is widely
accepted that giant planet cores must form via rapid runaway growth
\citep{Wetherill1989} and oligarchic growth \citep{Kokubo1998}, which
can grow the cores for the giant planet (5--10\,$M_\earth$) at the outer
disk (5\,AU) in a reasonable time, e.g, 10\,Myr. Because this growth
time is comparable to $\tau_{\rm disk}$, the variation of $\tau_{\rm
disk}$ due to the metallicity around this timescale may cause a strong
metallicity dependence of the planet formation probability, $P_{\rm
pl}$, thus contributing to the planet-metallicity correlation.

As far as we know, a deterministic model based on the core-accretion
model by Ida \& Lin (2004) is the only study that quantitatively
estimates the effect of $\tau_{\rm disk}$ variation on $P_{\rm
pl}$. They show that the planet formation probability decreases with
decreasing $\tau_{\rm disk}$ (see their Fig.2 (b)).  However, the amount
of decrease is only a factor of 2--4 with a factor of 10 change in the
disk lifetime: the dependence is more like $P_{pl} \sim {\tau_{\rm
disk}^{0.5}}$ and is not a strong function of $\tau_{\rm disk}$. If this
is the case, our suggestion that $\tau_{\rm disk} \sim 10^Z$ means the
disk dispersal can contribute to the planet-metallicity correlation by a
factor of $10^{0.5Z}$. \cite{Ida2004ApJ616} show that the core-accretion
model can qualitatively explain the planet-metallicity relation, but
also note that the suggested metallicity dependence is not as strong as
the observed relation: it is more like $\sim10^Z$ dependence judging
from their Fig.~2 (b). Therefore, by adding the contribution from the
disk dispersal ($\sim10^{0.5Z}$), most of the observed metallicity
dependence of the planet formation probability ($10^{2Z}$) can be
reasonably explained. However, a quantitative argument is difficult at
this stage because there are ambiguities and complexities in the planet
formation models, and thus the planet-formation time scale.  Note that
other groups are trying to reproduce the steep metallicity dependence by
introducing a critical solid mass \citep{Wyatt2007,Matsuo2007} for the
core accretion model.


While the planet-metallicity correlation appears to be clear in
Fischer's systematic work \citep{Fischer2005}, Udry \& Santos suggest
that the correlation in low-metallicity range ([Fe/H] $< 0$) may be
relatively weak \citep{Udry2007}.  Santos suggests that the planet
formation mechanism could be different for the two metallicity ranges:
core-accretion for the high-metallicity range and disk instability for
the low-metallicity range \citep{Santos2008}.
Because our results seem to qualitatively explain a significant part of
the 10$^{2Z}$ dependence of the planet-metallicity relation, we assume
that the dependence also holds true in the low-metallicity range ([Fe/H]
$<$ 0), where we made our observations.
However, the core-accretion model may not be able to explain the
existence of planets around the low-metallicity stars with such short
disk lifetime.


\section{Summary}

We present deep imaging of two low-metallicity star forming regions in
the EOG, Digel Cloud 2-N and Cloud 2-S.

\begin{enumerate}

\item The fraction of young stars with disks is found to be 9$\pm$4\,\%
      for Cloud 2-N and 27$\pm$7\,\% for Cloud 2-S. Previous works have
      shown that the age of these clusters is only approximately
      0.5\,Myr, and therefore the disk fraction is much lower than
      expected compared to the disk fraction measured for nearby star
      forming regions.

\item Various possibilities for the reason for the low disk fraction are
      investigated. We suggest that the simple reduction of the
      dust-to-gas ratio cannot explain the low disk fraction and that
      the initial disk fraction should be as high as in the solar
      neighborhood. We also suggest that dust sublimation and stellar
      magnetic activity cannot explain the enlargement of the inner disk
      hole in the low-metallicity environment.  Therefore, the
      significant decrease of the disk fraction implies that the disk
      lifetime is much shorter ($\sim$1\,Myr) than in the solar
      neighborhood (5--6\,Myr). We suggest that the short disk lifetime
      is a result of the photoevaporation of the surface layer of the
      gas disk.

\item We suggest that the lower disk lifetime with lower metallicity may
      be a major factor in explaining the planet-metallicity
      correlation. But this is model-dependent and further theoretical
      and observational work is needed.

\end{enumerate}

Although we derived disk fractions for two very young clusters in this
paper, it is necessary to study more embedded clusters in the EOG to
find a more quantitative relationship between the disk lifetime and the
metallicity. It is also important to study the disks in longer
wavelengths, e.g., $L$-band and mid-infrared wavelength, to confirm if
disks at larger radii also disappear rapidly, as suggested. These
results will be a strong constraint for interpreting the
planet-metallicity correlation, and ultimately for understanding how
many planet-harboring stars we can expect in the Galaxy
\citep{Lineweaver2004}.


\acknowledgments

C.Y has been supported by the Japan Society for the Promotion of
Science (JSPS).
We thank Dr. Ichi Tanaka for assistance with the observation and data
reduction.
We also thank Dr. Taku Takeuchi, Dr. Takeru Suzuki, Dr. Takayoshi Sano,
and Dr. Frank Shu for helpful discussions on theoretical issues.
We also thank the anonymous referee for the careful reading and
thoughtful suggestions that improved this paper significantly.


\appendix

\section{Comparison of Disk Fractions in Different Filter Systems for
the NGC 2024 Cluster}

NGC 2024 is a good reference target for the comparison of disk fractions
in different filter systems because it is a very young (0.3\,Myr)
embedded cluster in the solar neighborhood \citep{Meyer1997} with a high
disk fraction \citep{Lada1999}.  The data were obtained on 1 January
2007 (based on data collected at the Subaru Telescope and obtained from
the SMOKA, which is operated by the Astronomy Data Center, National
Astronomical Observatory of Japan) with a field of view of $\sim$$3'
\times 4'$ centered on $(\alpha_{\rm 2000}, \beta_{\rm 2000}) =
(05:41:41.4, -01:54:59.84)$.  The NGC 2024 data were reduced using the
same procedures as the Cloud 2 data.  For the photometry, we used only
stars that are sufficiently isolated with an aperture radius of
$d=3.4''$ ($\sim$95\,\% energy).  For the photometric calibration, we
used five stars in the field, whose $JHK$ photometric quality in 2MASS
is good, after converting 2MASS magnitudes to MKO magnitudes.

Disk fractions can change in different mass ranges and in different
spatial regions on the sky even when observed in the same filter system.
The disk fraction estimated with stars, which are in the same region on
the sky as the MOIRCS data and in almost the same mass range of
$\sim$0.7--0.02 $M_\odot$ from a previous study in the CIT system
\citep{Levine2006}, was 55$\pm$17\,\% (10/18), which is consistent with
the MOIRCS data (65$\pm$16\,\%, 17/26) within the errors.  Moreover,
almost all stars in the disk excess region in the CIT color-color
diagram are located in the same excess region in the MKO color-color
diagram. This is true also for the stars in the non-disk excess region.
Therefore, we conclude that the disk fraction of NGC 2024 is correctly
estimated in the MKO system.



\vspace{-25mm}

\begin{table}[h]
\begin{center}
\caption{$JHK$ disk fractions of embedded clusters in the solar
 neighborhood.}
\begin{tabular}{ccccccc}
\tableline\tableline
ID & Cluster & Age\tablenotemark{a} & $JHK$ disk fraction &
 Mass\tablenotemark{b} &
 Filter system\tablenotemark{c} & Reference\tablenotemark{d}\\ 
 & & (Myr) & (\%) &  ($M_\odot$) & \\
\hline
1 & NGC 2024 
& 0.3 (1) 
& 58$\pm$7
& 0.13 
& SQUIID
& 2 
\\
2 &  Trapezium 
& 1.5 (3) 
& 53$\pm$3 
& $\sim$0.03
& SQUIID
& 4 
\\
3 &  Cham I 
& 1.8 (3) 
& 41$\pm$4 
& $\sim$0.03 
& CIT
& 5 
\\
4 &  Taurus 
& 2.3 (3) 
& 34$\pm$6 
& 0.3 
& SQUIID
& 6 
\\
5 &  IC 348 
& 2.3 (3) 
& 21$\pm$4 
& 0.19 
& SQUIID
& 7 
 \\
6 & NGC 2264 
& 3.3 (3) 
& 30$\pm$3
& $\sim$0.3
& SQUIID
& 8 
\\
7 & Tr 37
& 4.0 (9) 
& 17$\pm$3 
& $\sim$0.3
& 2MASS
& 9 
\\
\hline
\end{tabular}
\end{center}
\tablenotetext{a}{References for the ages are shown in the parenthesis.}
\tablenotetext{b}{Mass detection limit of the data.}
\tablenotetext{c}{The photometric system of the $JHK$ data obtained with
the Simultaneous Quad Infrared Imaging Device (SQUIID) at Kitt Peak
National Observatory is similar to the Johnson system and the CIT
systems \citep{{Lada1993},{Horner1997}}}
\tablenotetext{d}{References for the $JHK$ photometric data. For
Trapezium, only the Fred Lawrence Whipple Observatory (FLWO) data were
used to match the mass detection limit \citep[see][]{Muench2002}.  A
$K_S$ filter was used instead of $K$ filter for the observation of Tr
37.}
\tablerefs{(1) \citealt{Meyer1997}; (2) \citealt{Haisch2000AJ}; (3)
\citealt{Palla2000}; (4) \citealt{Muench2002}; (5) \citealt{Lhuman2004};
(6) \citealt{Kenyon1995}; (7) \citealt{Haisch2001AJ}; (8)
\citealt{Rebull2002}; (9) \citealt{Sicilia-Aguilar2005ApJ130}}
\end{table}

\end{document}